\documentclass[10pt, doublecolumn]{IEEEtran}
\usepackage{graphicx}
\usepackage{amsmath}
\usepackage{amsfonts}
\usepackage{amssymb}

\begin{document}

\title{Interference Channel with a Half-Duplex Out-of-Band Relay }
\author{Onur Sahin$^*$, Osvaldo Simeone$^\dag$, and Elza Erkip$^\ddag$ \\
%EndAName
* InterDigital Inc., New York\\
\dag CWCSPR, ECE\ Dept, NJIT\\
$\ddag$Dept. of ECE, Polytechnic Inst. of NYU\\
Email: onur.sahin@interdigital.com, osvaldo.simeone@njit.edu, elza@poly.edu
\thanks{%
This work is partially supported by U.S. NSF grants No. 0520054, 0635177 and
0905446 and The Wireless Internet Center for Advanced Technology at
Polytechnic Institute of NYU. The work of O. Simeone is supported by U.S.
NSF grant CCF-0914899.}}
\date{}
\maketitle

\begin{abstract}
A Gaussian interference channel (IC) aided by a half-duplex relay is
considered, in which the relay receives and transmits in an orthogonal band
with respect to the IC. The system thus consists of two parallel channels,
the IC and the channel over which the relay is active, which is referred to as Out-of-Band Relay Channel (OBRC). The OBRC is operated by separating a
multiple access phase from the sources to the relay and a broadcast phase from the
relay to the destinations. Conditions under which the optimal operation, in terms of the
sum-capacity, entails either \textit{signal relaying} and/or \textit{%
interference forwarding} by the relay are identified. These conditions also
assess the optimality of either separable or non-separable transmission over
the IC and OBRC. Specifically, the optimality of signal relaying and
separable coding is established for scenarios where the relay-to-destination
channels set the performance bottleneck with respect to the source-to-relay
channels on the OBRC. Optimality of interference forwarding and
non-separable operation is also established in special cases.
\end{abstract}

% Declares the document's title.

%The aim of this work is to give a brief introduction to the system
%shown in Figure \ref{Fig:System}. We investigate a network that
%consists of an interference channel and a relay with orthogonal and
%finite-capacity links. Achievable schemes as well as rate regions
%are mentioned. Intuitive upper bounds follow where in special cases
%lead to capacity results.

\section{Introduction}

Consider two interfering links, say belonging a Wireless Local Area Network,
that operate over the same bandwidth, i.e., an interference channel (IC).
The corresponding transmitters and receivers of the IC may be also endowed
with a second, shorter-range, radio interface, such as Bluetooth, that can
be used for communications with an external terminal over an orthogonal
bandwidth, as shown in Fig. 1. This terminal may act as a relay for both
links, while operating out-of-band with respect to the IC\ (i.e., as an
Out-of-Band Relay, or OBR). By this means, communication takes place
effectively over two parallel channels, the IC and the channel where the
relay is active, which is termed as OBR channel (OBRC). We refer to the
overall channel comprising IC and OBRC as IC-OBR. This scenario was first
considered in \cite{onur_ISIT09}, where it was assumed that the OBRC\ is
operated via an orthogonal medium access scheme (e.g., TDMA) that makes the
links from each transmitter to the relay, and from the relay to each
receiver, all orthogonal to one another. In this paper, we study the more
complex situation in which the relay is simply assumed to be half-duplex, so
that the OBRC is operated by allowing the relay to either transmit or
receive at a given time.

The considered model is related to, and inspired by, two recent lines of
work. The first deals with \textit{relaying in interference-limited systems}%
, where, unlike in the IC-OBR, the relay is assumed to operate in the same
band as the IC \cite{onur_globecom}-\cite{Aylin Globecom09}\cite{onur_ISIT09}.
These works reveal the fact that relaying in interference-limited systems
offers performance benefits not only due to \emph{signal relaying,} as for
standard relay channels, but also thanks to the novel idea of \emph{%
interference forwarding}. According to the latter, the relay may help by
reinforcing the interference received at the undesired destination, so as to
facilitate interference stripping, exploiting the standard technique of rate
splitting into \textit{private and common} messages (see, e.g., \cite{Raul}%
): Private messages are decoded only at the intended destination, while
common messages are decoded also at the interfered destination (and may
benefit from interference forwarding).

A second related line of work deals with communications over \textit{%
parallel ICs }(albeit the considered OBRC\ is not a conventional IC). As
shown in \cite{cadambe}-\cite{Koreans}, optimal operation over parallel ICs,
unlike scenarios with a single source or destination, typically entails
\textit{joint}, rather than \textit{separate}, coding over the parallel
channels. In other words, the signals sent over the parallel ICs need to be
generally correlated to achieve optimal performance. The question arises as
to what type of information, either private or common, should be sent in a
correlated fashion over the component ICs. For instance, the original work
\cite{cadambe} derives conditions under which correlated transmission of
private messages is optimal, \cite{Lalitha} considers the optimality of
common information transmission, whereas in \cite{Koreans} scenarios are
found for which sending both correlated private and common messages is
optimal.

In this paper, we study the IC-OBR model and derive conditions under
which a separable coding scheme with only signal relaying is sum-rate
optimal, and also conditions under which a non-separable coding scheme with
both signal relaying and interference forwarding achieve optimal performance.
Analytical results are corroborated by numerical examples.
\begin{figure}[t]
\begin{center}
\includegraphics[width=7 cm]{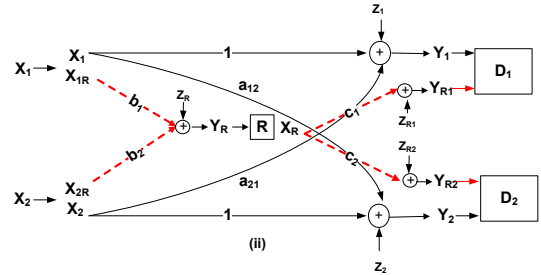}
\end{center}
\caption{Interference Channel (IC) with an out-of-band relay (OBR). }
\label{Fig:System}
\end{figure}

\section{System Model}

\label{Sec:Type-II}

The Gaussian IC-OBR\ model, shown in Fig. \ref{Fig:System}, consists of two parallel
channels, namely the IC and the OBRC. On the IC, the signal received by
destination $D_{i}$ at time $t=1,...,n,$ is given by%
\begin{equation}
Y_{i,t}=X_{i,t}+a_{ji}X_{j,t}+Z_{i,t},  \label{Y i t}
\end{equation}%
where $i,j=1,2$ and $i\neq j,$ $Z_{i,t}$ are independent unit-power white
Gaussian noise sequences, and $X_{i,t}$ are the transmitted sequences over
the IC. Due to the half-duplex constraint, the OBRC is orthogonalized into
two channels, one being a multiple-access channel (MAC) from $S_{1}$ and $%
S_{2}$ to $R,$ with fraction of channel uses $\eta _{MAC}$ (relay listening)$%
,$ and the other being a broadcast channel (BC) from $R$ to $D_{1}$ and $%
D_{2}$, with fraction of channel uses $\eta _{BC}$ (relay transmitting). We
have $\eta _{MAC}+\eta _{BC}=\eta ,$ where $\eta $ is the ratio between the
bandwidths (and thus the channel uses) of the OBRC\ and of the IC. The
received signal at the relay $R$ over the OBRC\ is given by the MAC
relationship
\begin{equation}
Y_{R,t}=b_{1}X_{1R,t}+b_{2}X_{2R,t}+Z_{R,t}  \label{Y R t}
\end{equation}%
for $t=1,...,\eta _{MAC}n$ \footnote{%
We will not denote explicitly the necessary integer rounding-off operation.}%
; and the signal received at destination $D_{i}$ over the OBRC is given by
the BC relationships
\begin{equation}
Y_{Ri,t}=c_{i}X_{R,t}+Z_{Ri,t},  \label{Y Ri t}
\end{equation}%
for $t=1,...,\eta _{BC}n$, and $i=1,2.$ We have the power constraints $\frac{%
1}{n}\sum_{t=1}^{n}E[X_{i,t}^{2}]\leq P_{i}$ on the IC, and $\frac{1}{n}%
\sum_{t=1}^{\eta _{MAC}n}E[X_{iR,t}^{2}]\leq P_{iR}$, $\frac{1}{n}%
\sum_{t=1}^{\eta _{BC}n}E[X_{R,t}^{2}]\leq P_{R}$ on the OBRC, $i=1,2$.
Bandwidth allocations ($\eta _{MAC},\eta _{BC}$) will be considered as fixed
and given throughout the paper, except in Sec. \ref{Sec:Var_BW}.

Encoding and decoding functions, probability of error and achievable rates
are defined in the usual way. In particular, encoding at the source $S_{i}$
produces two sequences\footnote{$X^{n}\triangleq (X_{1},...,X_{n}).$}, one
transmitted on the IC, $X_{i}^{n}$, and one on the OBRC, $X_{iR}^{n}$. The
relay decides the transmitted codeword $X_{R}^{\eta _{BC}n}$ based on the
signal received from the sources, namely $Y_{R}^{\eta _{MAC}n}$ (\ref{Y R t})%
$.$ Finally, the destination $D_{i}$ decodes on the basis of the signals
received over the IC, $Y_{i}^{n}$\ (\ref{Y i t}), and over the OBRC, $%
Y_{Ri}^{\eta _{BC}n}$ (\ref{Y Ri t}).

\section{Scenario and Achievable Strategies\label{sec_scenario}}

The model at hand appears too complicated to hope for general conclusions
regarding the capacity region. In fact, the sources may employ a number of
rate splitting strategies, with independent or correlated transmission of
information over the two parallel channels, IC and OBRC, and may deploy
either structured or unstructured codes. Moreover, the relay may implement a
number of relaying strategies, encompassing regenerative or non-regenerative
techniques. It is emphasized that, not only the optimal operation on the IC
alone is generally not known \cite{Raul}, but the same also holds for the
operation over the OBRC model alone\footnote{%
A special case of this model is the two-way relay channel, whose capacity is
still generally unknown. Note that the OBRC model is significantly more complex than the four orthogonal links considered in \cite{onur_ISIT09}.}. Therefore, in this paper, we focus on specific
channel gain conditions and suitable achievable strategies. We will show
optimality of the considered techniques in a number of special cases of
interest.

In particular, we will consider a scenario in which interference towards
receiver $D_{2}$ is weak, i.e., $a_{12}\leq 1$ and the channel from the
relay to destination $D_{2}$ is worse than towards destination $D_{1},$
i.e., $c_{1}\geq c_{2}$. Under these conditions, we consider the performance
of strategies whereby transmitter $S_{1},$ interfering on $D_{2}$ transmits:
(\textit{i}) on the IC only \textit{private} information, which is then
treated as noise by $D_{2};$ (\textit{ii}) on the OBRC independent
information, thus using \textit{separate} encoding over IC\ and OBRC. This
choice appears to be reasonable in light of the channel conditions mentioned
above. In contrast, transmitter $S_{2}$, whose interference may not
necessarily be weak, transmits: (\textit{i}) on the IC with \textit{both
private and common} messages; (\textit{ii}) on the OBRC with the \textit{%
same common message} plus an additional independent message. Transmitter $%
S_{2}$ thus potentially employs a non-separable coding strategy where the
same (common) message is sent over both IC and OBRC. Since this message is
common, the operation of the OBR can be classified as \textit{interference
forwarding} \cite{onur_ISIT09}\cite{dabora_ITW08}: The relay forwards information about the
interference on the IC from $S_{2}$ to $D_{1}$. Moreover, we assume that the
OBR employs Decode-and-Foward (DF). The main questions of interest are:
Under what conditions is the scheme described above optimal? And, when this
is the case, under what conditions is separable (rather than the general
non-separable) coding at transmitter $S_{2}$ optimal?

\section{Outer Bounds}

\label{Sec:IC-OBR Type II} In this section, we give a general outer bound on
the capacity region of IC-OBR.

\textbf{Proposition 1 (Outer Bound for IC-OBR):} For an IC-OBR for $%
a_{12}\leq 1$ and $c_{1}\geq c_{2},$ with given bandwidth allocation ($\eta
_{MAC},\eta _{BC}$), the capacity region is included in the following region
\begin{subequations}
\label{R2outc_end}
\begin{align}
& \lim_{n\rightarrow \infty }\mbox{closure}\bigg(%
\bigcup_{p(x_{1}^{n},x_{2}^{n})=p(x_{1}^{n})p(x_{2}^{n}),0\leq \xi +%
\overline{\xi }\leq 1}\big\{(R_{1},R_{2})\text{:}  \notag \\
R_{j}& \leq \frac{1}{n}I(X_{j}^{n};Y_{j}^{n})+\eta _{MAC}\mathcal{C}\left(
b_{1}^{2}P_{1R}+b_{2}^{2}P_{2R}\right) ,\text{ }  \label{R2outc_end_b1} \\
R_{j}& \leq \frac{1}{n}I(X_{j}^{n};Y_{j}^{n}|X_{i}^{n})+\eta _{MAC}\mathcal{C%
}\left( b_{j}^{2}P_{jR}\right) ,\text{ }j=1,2,\text{ }i\neq j
\label{R2outc_end_b2} \\
R_{1}& \leq \frac{1}{n}I(X_{1}^{n};Y_{1}^{n}|X_{2}^{n})+\eta _{BC}\mathcal{C}%
(c_{1}^{2}\xi P_{R})  \label{R2outc_end_b3} \\
R_{2}& \leq \frac{1}{n}I(X_{2}^{n};Y_{2}^{n})+\eta _{BC}\mathcal{C}\left(
\frac{c_{2}^{2}\overline{\xi }P_{R}}{1+c_{2}^{2}\xi P_{R}}\right) \big\}%
\bigg),  \label{R2outc_end_b4}
\end{align}%
where the union is taken with respect to multi-letter input distributions $%
p(x_{1}^{n})p(x_{2}^{n})$ that satisfy the power constraints $%
1/n\sum_{t=1}^{n}\mbox{E}[X_{i,t}^{2}]\leq P_{i}$, $i=1,2$, and with respect
to parameters $\xi ,$ with $0\leq \xi \leq 1$ and $\overline{\xi }=1-\xi .$

\emph{Proof:} Appendix A.

\section{Capacity Results}

In this section, we consider fixed OBRC bandwidth allocation ($\eta
_{MAC},\eta _{BC}$) and derive two sets of conditions for optimal operation.
Under the first, a special case of the strategy described in Sec. \ref%
{sec_scenario}, in which separable coding only (at both transmitters) is
employed, is shown to be optimal, while the second set of conditions
provides (asymptotic) optimality of the general non-separable technique.

\subsubsection{Optimality of Separable Encoding}
%\\*
\[\]

\vspace{-0.35in}

\textbf{Proposition 2:} In an IC-OBR with $a_{12}\leq 1$, $c_{1}\geq c_{2}$
\textit{and} $a_{21}\geq \sqrt{\frac{1+P_{1}}{1+a_{12}^{2}P_{1}}}$, the
sum-capacity is given by
\end{subequations}
\begin{equation}
R_{1}+R_{2}\leq \mathcal{C}\left( P_{1}\right) +\mathcal{C}\left( \frac{P_{2}%
}{1+a_{12}^{2}P_{1}}\right) +\eta _{BC}\mathcal{C}\left(
c_{1}^{2}P_{R}\right)  \label{Prop11ratesum1}
\end{equation}%
if condition $\eta _{MAC}\mathcal{C}(b_{1}^{2}P_{1R})\geq \eta _{BC}\mathcal{%
C}(c_{1}^{2}P_{R})$ holds; and by
\begin{align}
R_{1}+R_{2}& \leq \max_{0\leq \xi \leq 1}\bigg\{\mathcal{C}\left(
P_{1}\right) +\mathcal{C}\left( \frac{P_{2}}{1+a_{12}^{2}P_{1}}\right)
\notag \\
& +\min \big\{\eta _{MAC}\mathcal{C}\left( b_{1}^{2}P_{1R}\right) ,\eta _{BC}%
\mathcal{C}\left( c_{1}^{2}\xi P_{R}\right) \big\}  \notag \\
& +\eta _{BC}\mathcal{C}\left( \frac{c_{2}^{2}\overline{\xi }P_{R}}{%
1+c_{2}^{2}\xi P_{R}}\right) \bigg\},  \label{Prop11ratesum2}
\end{align}%
if conditions $\eta _{MAC}\mathcal{C}(b_{1}^{2}P_{1R})<\eta _{BC}\mathcal{C}%
(c_{1}^{2}P_{R})$ and $\eta _{MAC}\mathcal{C}\left( \frac{b_{2}^{2}P_{2R}}{%
1+b_{1}^{2}P_{1R}}\right) \geq \eta _{BC}C\left( \frac{c_{2}^{2}\overline{%
\xi }^{\ast }P_{R}}{1+c_{2}^{2}\xi ^{\ast }P_{R}}\right) $ hold, where $\xi
^{\ast }$ is the optimal power allocation that maximizes the sum-rate (\ref%
{Prop11ratesum2}) with $\overline{\xi }^{\ast }=1-\xi ^{\ast }$. The sum-capacity is achieved by separable coding on IC and OBRC.

\emph{Proof:} The converse follows from Proposition 1 and invoking the
worst-case noise result of \cite{worst-case} applied for $a_{12}\leq 1$. For
the achievability, we follow the strategy described in Sec. \ref%
{sec_scenario} and we refer to Appendix B for further details.$\Box $
\begin{figure}[t]
\begin{center}
\includegraphics[width=7 cm]{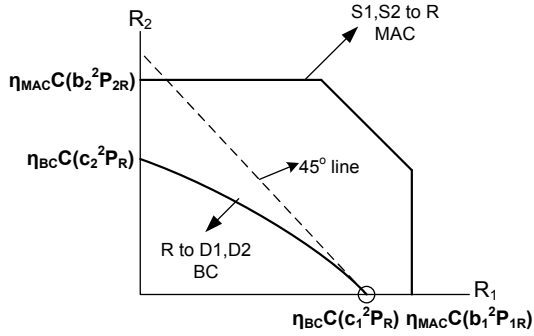}
\end{center}
\caption{Illustration of the first OBRC condition leading to the
sum-capacity in Proposition 2.}
\label{Fig:MAC_BC_a}
\end{figure}
\begin{figure}[t]
\begin{center}
\includegraphics[width=7 cm]{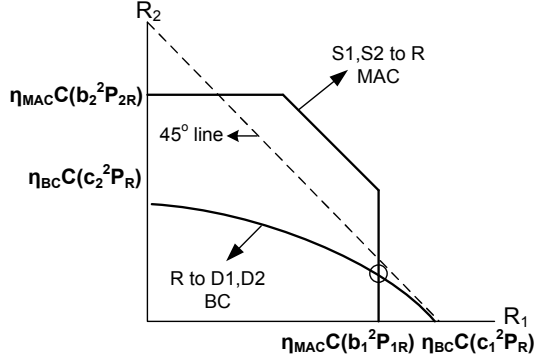}
\end{center}
\caption{Illustration of the second OBRC condition leading to the
sum-capacity in Proposition 2.}
\label{Fig:MAC_BC_b}
\end{figure}

Optimality in Proposition 2 is achieved by using a special case of the
strategy described in Sec. \ref{sec_scenario} in which transmitter 1
operates as prescribed, and transmitter 2 sends only common information on
the IC and transmits independent information over the OBRC\ (separable
coding). Notice that transmission of only common information over the IC\ is
justified by the``strong interference'' condition $a_{21}\geq \sqrt{%
(1+P_{1})/(1+a_{12}^{2}P_{1})}.$ Also, notice that here the relay performs
only signal relaying \cite{onur_ISIT09}. To be more specific, we have two
subcases depending on the channel gains of the OBRC.

The first set of conditions (under which the sum-capacity is (\ref%
{Prop11ratesum1})) is characterized by $\eta _{MAC}\mathcal{C}%
(b_{1}^{2}P_{1R})\geq \eta _{BC}\mathcal{C}(c_{1}^{2}P_{R})$ and is
illustrated in Fig. \ref{Fig:MAC_BC_a}. It corresponds to the case where the
relay-to-destinations BC constitutes the \emph{bottleneck} with respect to
the sources-to-relay MAC in the $S_{1}-R-D_{1}$ communication path. In this
case, the optimal strategy in terms of sum-rate is for user 1 only to
transmit over the OBRC. Notice that this operating point on the OBRC\ (see
dot in the figure) is sum-rate optimal if one focuses on the OBRC alone
limiting the scope to DF techniques, since the corresponding achievable rate
region is given by the intersection of the MAC and BC\ regions in Fig. \ref%
{Fig:MAC_BC_a}. Proposition 2 shows that such operating point is also
optimal for communications over the IC-OBR under the given conditions.\

The second set of conditions (under which the sum-capacity is (\ref%
{Prop11ratesum2})) is given by $\eta _{MAC}\mathcal{C}(b_{1}^{2}P_{1R})<\eta
_{BC}\mathcal{C}(c_{1}^{2}P_{R})$ and $\eta _{MAC}\mathcal{C}\left( \frac{%
b_{2}^{2}P_{2R}}{1+b_{1}^{2}P_{1R}}\right) \geq \eta _{BC}C\left( \frac{%
c_{2}^{2}\overline{\xi }^{\ast }P_{R}}{1+c_{2}^{2}\xi ^{\ast }P_{R}}\right) $%
, and is illustrated in Fig. \ref{Fig:MAC_BC_b}. Here, the sum-rate optimal
operation of the OBRC entails transmission over the OBRC (of independent
information) by both sources, not just source 1 as above, at rates given by
the operating point indicated in the figure. This rate pair is characterized
by the power split $\xi ^{\ast }$ in Proposition 2$.$ Notice that the same
considerations given above regarding optimality of this point for the OBRC\
alone with DF apply here.

%\textbf{Remark 12:} Observing Fig. \ref{Fig:MAC_BC_b} and recalling
%Propositions 4-6, one could guess (erroneously, as Proposition 2 shows) that
%signal relaying is suboptimal under the conditions of Fig. \ref{Fig:MAC_BC_b}. These conditions seem to suggest an ``excess rate'', in the sense of
%Proposition 4 and Remark 5, between $S_2$ and $D_1$, since the
%maximum rate $S_{2}-R$ (i.e., $\eta_{MAC}\mathcal{C}(b_{2}^{2}P_{2R})$) is
%larger than the maximum rate $R-D_{2}$ (i.e., $\eta_{BC}\mathcal{C}(c_{2}%
%^{2}P_{R})$), while the opposite is true for the path $S_{1}-R-D_{1}.$ The
%fact that such excess rate is not to be exploited for interference forwarding
%by the optimal scheme of Proposition 2 can be interpreted in light of Remark
%8, since the condition $a_{21}\geq\sqrt{\frac{1+P_{1}}{1+a_{12}^{2}P_{1}}},$
%assumed in Proposition 2, already guarantees strong interference condition at
%$D_{1},$ and thus no need for interference forwarding. For $a_{21}%
%<\sqrt{\frac{1+P_{1}}{1+a_{12}^{2}P_{1}}}$, we will show that
%interference-forwarding is instead instrumental in maximizing the sum-rate,
%see Sec. \ref{Subsec:Type_II_Int_Forw}.

\subsubsection{Optimality of Non-Separable Encoding\label%
{Subsec:Type_II_Int_Forw}}

In this section, we discuss conditions under which strategies based on
interference forwarding, and thus non-separable encoding, are optimal. In
Proposition 2, the ``strong interference'' condition $a_{21}\geq \sqrt{%
(1+P_{1})/(1+a_{12}^{2}P_{1})}$ at $D_{1}$ was instrumental in making
interference forwarding from $S_{2}$ to $D_{1}$ not necessary. Consider then
a more general situation in which this condition is not imposed. We will see
below that, in this case, interference forwarding (and non-separable coding)
is potentially useful. This is shown by first providing an achievable region
for a special case of the scheme described in Sec. \ref{sec_scenario}, which
involves interference forwarding, then establishing its asymptotic
optimality under given conditions (that include the scenario discussed
above), and finally discussing some numerical results.

\textbf{Proposition 3}: In the IC-OBR, the following conditions
\begin{subequations}
\begin{align}
R_{1}& \leq \mathcal{C}(P_{1})+\eta _{MAC}\mathcal{C}(b_{1}^{2}P_{1R})
\label{Prop14R1ach} \\
R_{2}& \leq \mathcal{C}\left( \frac{P_{2}}{1+a_{12}^{2}P_{1}}\right) +\eta
_{BC}\mathcal{C}\left( \frac{c_{2}^{2}\overline{\xi }P_{R}}{1+c_{2}^{2}\xi
P_{R}}\right) \\
R_{1}+R_{2}& \leq \mathcal{C}(P_{1}+a_{21}^{2}P_{2})+\eta _{BC}\mathcal{C}%
(c_{1}^{2}\xi P_{R})\nonumber\\&+\eta _{BC}\mathcal{C}\left( \frac{c_{2}^{2}\overline{%
\xi }P_{R}}{1+c_{2}^{2}\xi P_{R}}\right) \\
R_{1}+R_{2}& \leq \mathcal{C}(P_{1}+a_{21}^{2}P_{2})+\eta _{MAC}\mathcal{C}%
(b_{1}^{2}P_{1R}+b_{2}^{2}P_{2R})  \label{Prop14Rtotach}
\end{align}%
with $\xi +\overline{\xi }\leq 1,$ define a rate region achievable with the
scheme of Sec. \ref{sec_scenario}. Moreover, for $a_{12}\leq 1$, and $%
b_{2},c_{1}\rightarrow \infty $, the sum-capacity is achieved by this scheme
and given by
\end{subequations}
\begin{eqnarray}
R_{1}+R_{2}&&\leq \mathcal{C}(P_{1})+\mathcal{C}\left( \frac{P_{2}}{%
1+a_{12}^{2}P_{1}}\right) +\eta _{MAC}\mathcal{C}(b_{1}^{2}P_{1R})\nonumber\\&&+\eta _{BC}%
\mathcal{C}(c_{2}^{2}P_{R}).
\end{eqnarray}

\emph{Proof:} Appendix B.

The scheme achieving (\ref{Prop14R1ach})-(\ref{Prop14Rtotach}) is based on $%
S_{2}$ transmitting common information over the IC, as for Proposition 2,
and both the same common and an independent private message on the OBRC\
(non-separable encoding). This scheme is shown in Proposition 3 to be
sum-rate optimal if a very good channel is available between $S_{2}$ and $%
D_{1}$ through the relay, so as to essentially drive $D_{1}$ back in the
``strong interference regime'' thanks to interference forwarding. It is noted
that this condition is related to the ``large excess rate'' assumption of
Theorem 4 in \cite{onur_ISIT09} (which applies to a TDMA-based operation on
the OBRC).

\begin{figure}[t]
\begin{center}
\includegraphics[width=8 cm]{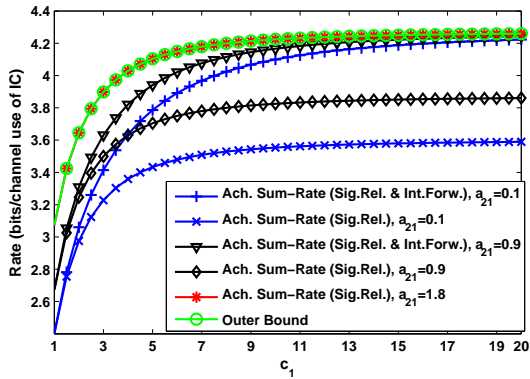}
\end{center}
\caption{Achievable sum-rate and outer bound for an IC-OBR with respect to $%
R-D_{1}$ channel gain, $c_{1}$ and $S_{2}-D_{1}$ channel gain $a_{21}\in
\{0.1,0.9.1.8\}$ ($a_{12}=0.5$, $b_{1}=1$, $b_{2}=10$, $c_{2}=1$ and all
node powers are equal to 10).}
\label{Fig:ICOBRII_Int_forw}
\end{figure}

To investigate the role of interference forwarding in a non-asymptotic
regime, Fig. \ref{Fig:ICOBRII_Int_forw} shows the sum-rate obtained from (%
\ref{Prop14R1ach})-(\ref{Prop14Rtotach}), by assuming that source 2 either
transmits only an independent message on the OBRC\ (signal relaying, i.e., $%
R_{2c^{\prime }}=0$ in the achievable region given in Appendix B) or also
employs interference forwarding, and the sum-rate upper bound obtained from
Proposition 1. The OBRC gains are set to $b_{1}=1$, $b_{2}=10$, $c_{2}=1$
and $c_{1}$ is varied, all node powers are equal to 10 and $\eta _{MAC}=\eta
_{BC}=1$. We also have $a_{12}=0.5$ and $a_{21}\in \{0.1,0.9.1.8\}$. Note
that for $a_{21}=1.8\geq \sqrt{(1+P_{1})/(1+a_{12}^{2}P_{1})}=1.78$ the
conditions given in Proposition 2 are satisfied and signal relaying alone is
optimal. For $a_{21}\in \{0.1,0.9\}\leq 1.78,$ instead, the advantages of
interference forwarding become substantial with increasing $c_{1}$, which is
due to the fact that the $S_{2}-D_{1}$ pair acquires an increasingly better
channel through the relay. Specifically, the asymptotic optimality derived
in Proposition 3 is here seen to be in practice attained for finite values
of $b_{2},c_{1}.$%

\section{Some Results for OBRC Variable Bandwidth Allocation}

\label{Sec:Var_BW}

\begin{figure}[t]
\begin{center}
\includegraphics[width=7 cm]{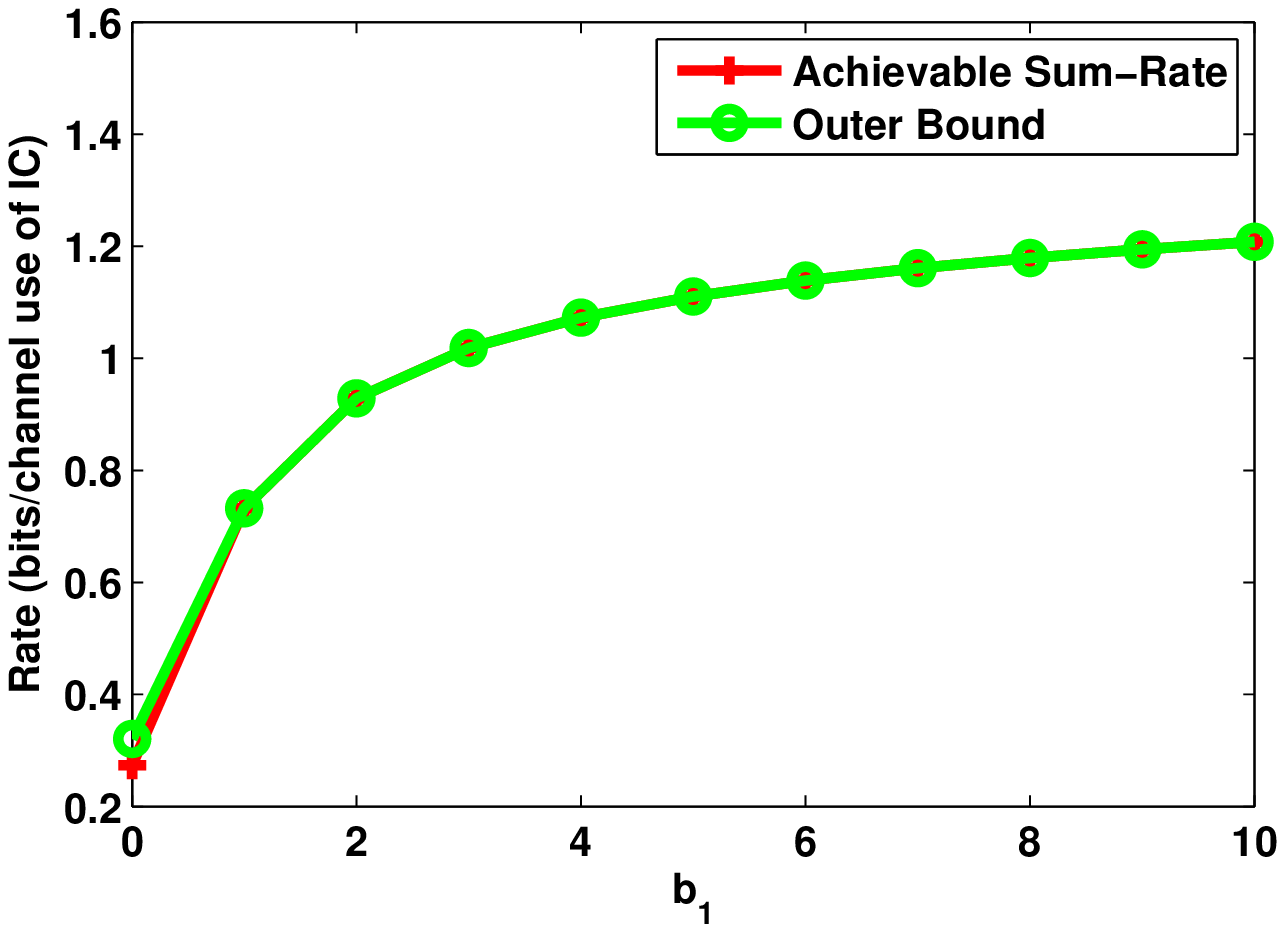} %
\includegraphics[width=7 cm]{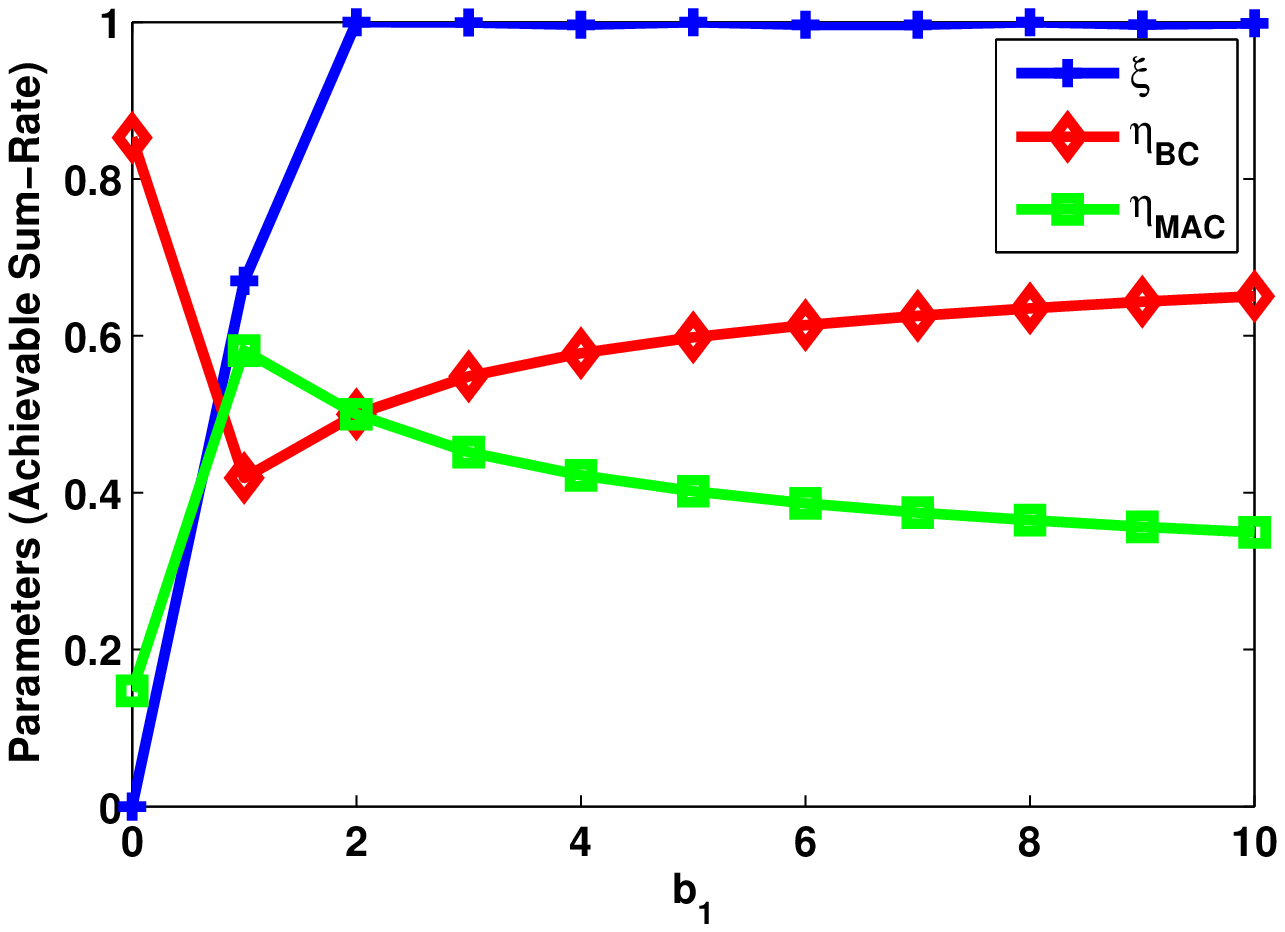}
\end{center}
\caption{ Achievable sum-rate (from Proposition 2, (\protect\ref%
{Prop11ratesum1})) with signal relaying and upper bound of Proposition 1 and
optimal parameters ($\protect\eta _{MAC}$, $\protect\eta _{BC}$, $\protect%
\xi $) for an IC-OBR with respect to $S_{1}-R$ channel gain $b_{1}$ ($%
b_{2}=2 $, $c_{1}=2$, $c_{2}=0.3$, $\protect\eta ={1}$, all node powers are
equal to $10$, $a_{21}=1.8$, $a_{12}=0.5).$ }
\label{Fig:Type_II_var_df_sing}
\end{figure}

We now investigate the effect of being able to optimize the bandwidth
allocation ($\eta _{MAC}$, $\eta _{BC}$) via numerical results. We consider
a scenario with $a_{12}=0.5$, $a_{21}=1.8$, $c_{1}\geq c_{2}$, and all
powers are set to $10$, which satisfies the conditions of Prop. 2,
except the ones that depend on the bandwidth allocation ($\eta _{MAC}$, $%
\eta _{BC}$), which is not specified a priori here. We compare the
performance of the achievable scheme of Prop. 2 (separable
transmission) with a sum-rate outer bound obtained from Prop. 1. In
both cases, the bandwidth allocation ($\eta _{MAC},\eta _{BC})$ is
optimized to maximize the sum-rate. In Fig. \ref{Fig:Type_II_var_df_sing} (upper part), the sum-rate
discussed above are shown for variable $S_{1}-R$ gain, $b_{1}$, and the
other channel gains are set to $b_{2}=2$, $c_{1}=2$, $c_{2}=0.3$ and $\eta
=1 $. We know from the first part of Prop. 2 that if $b_{1}$ is
sufficiently larger than $c_{1}$, for fixed bandwidth allocation, the rate (%
\ref{Prop11ratesum1}) where the relay helps the $S_{1}-D_{1}$ pair only, is
optimal. Observing the corresponding optimal bandwidth and power allocations for the achievable sum-rate
as shown in Fig. \ref{Fig:Type_II_var_df_sing}, a similar
conclusion is drawn here for $b_{1}\geq 2$ where the achievable sum-rate and outer bound coincide. Moreover, the total bandwidth is balanced between the $S_{1}-R$ and $R-D_{1}$ channels.

\section{Concluding Remarks}

\label{Sec:Conclusion}

Operation over parallel radio interfaces is bound to become increasingly
common in wireless networks due to the large number of multistandard
terminals. This enables cooperation among terminals across different
bandwidths and possibly standards. In this paper, we have studied one such
scenario where two source-destination pairs, interfering over a given
bandwidth, cooperate with a relay over an orthogonal spectral resource
(out-of-band relaying, OBR). We have derived analytical conditions under
which either signal relaying or interference forwarding are optimal. These
conditions have also been related to the problem of assessing optimality of
either separable or non-separable transmission over parallel interference
channels.

\section*{Appendix}

\section*{A. Proof of Proposition 1}

Bounds (\ref{R2outc_end_b2}) follow from cut-set arguments, while (\ref%
{R2outc_end_b1}) follows as
\begin{subequations}
\begin{align}
nR_{1}& \leq H(W_{1})  \label{Markov1} \\
& \leq I(W_{1};Y_{1}^{n},Y_{R1}^{\eta
_{BC}n})+H(W_{1}|Y_{1}^{n},Y_{R1}^{\eta _{BC}n}) \\
& \leq I(W_{1};Y_{1}^{n},Y_{R1}^{\eta _{BC}n})+n\epsilon _{n} \\
& \leq I(X_{1}^{n};Y_{1}^{n})+h(Y_{R}^{\eta _{MAC}n})-h(Z_{R}^{\eta
_{MAC}n})+n\epsilon _{n} \\
& \leq I(X_{1}^{n};Y_{1}^{n})+\eta _{MAC}n\mathcal{C}\left(
b_{1}^{2}P_{1R}+b_{2}^{2}P_{2R}\right) +n\epsilon _{n}
\end{align}%
from Fano's inequality, and from the Markov relations $W_{1}\rightarrow
Y_{R}^{\eta _{MAC}n},Y_{1}^{n}\rightarrow Y_{R1}^{\eta _{BC}n}$ and $%
W_{1}\rightarrow X_{1}^{n}\rightarrow Y_{1}^{n}$. Here $\epsilon _{n}\rightarrow 0$
as $n\rightarrow \infty$. We now focus on the
remaining two bound (\ref{R2outc_end_b3})-(\ref{R2outc_end_b4}), which
follow from considerations similar to the Gaussian BC. Proceeding as above,
we have
\end{subequations}
\begin{eqnarray}
nR_{2} &\leq &I(X_{2}^{n};Y_{2}^{n})+h(Y_{R2}^{\eta _{BC}n})  \notag \\
&&-h(c_{2}X_{R}^{\eta _{BC}n}+Z_{R2}^{\eta
_{BC}n}|Y_{2}^{n},W_{2})+n\epsilon _{n}.  \label{app R2}
\end{eqnarray}%
Now, consider the following
\begin{eqnarray*}
&&h(Z_{R2}^{\eta _{BC}n})\leq h(c_{2}X_{R}^{\eta _{BC}n}+Z_{R2}^{\eta
_{BC}n}|Y_{2}^{n},W_{2}) \\
&\leq &h(c_{2}X_{R}^{\eta _{BC}n}+Z_{R2}^{\eta _{BC}n})\leq \frac{\eta _{BC}n%
}{2}\log (2\pi e(1+c_{2}^{2}P_{R})),
\end{eqnarray*}%
so that, without loss of generality, one can define
\begin{equation*}
h(Y_{R2}^{\eta _{BC}n}|Y_{2}^{n},W_{2})=\frac{\eta _{BC}n}{2}\log (2\pi
e(1+c_{2}^{2}\xi P_{R})),
\end{equation*}%
for some $0\leq \xi \leq 1$. Then, (\ref{app R2}) becomes
\begin{equation*}
nR_{2}\leq I(X_{2}^{n};Y_{2}^{n})+\eta _{BC}n\mathcal{C}\left( \frac{%
c_{2}^{2}\overline{\xi }P_{R}}{1+c_{2}^{2}\xi P_{R}}\right) +n\epsilon _{n},
\end{equation*}%
where we have used the maximum entropy theorem. Now, consider
\begin{subequations}
\begin{eqnarray}  \label{scaling}
nR_{1}&\leq& I(W_{1};Y_{1}^{n},Y_{R1}^{\eta _{BC}n}|W_{2})+n\epsilon _{n}
\label{cond_dec_b} \\
&\leq& I(X_{1}^{n};Y_{1}^{n}|X_{2}^{n})+I(W_{1};Y_{R1}^{\eta
_{BC}n}|Y_{1}^{n},W_{2})+n\epsilon _{n} \\
&=&I(X_{1}^{n};Y_{1}^{n}|X_{2}^{n})+I(W_{1};\frac{c_{2}}{c_{1}}Y_{R1}^{\eta
_{BC}n}|Y_{1}^{n},W_{2})+n\epsilon _{n}.\nonumber\\
\end{eqnarray}
Since the capacity region of BC depends on the conditional marginal
distributions and noting that $c_{1}\geq c_{2}$, we can write $Y_{R2}^{\eta
_{BC}n}=\frac{c_{2}}{c_{1}}Y_{R1}^{\eta _{BC}n}+\widehat{Z}_{R}^{\eta
_{BC}n} $ where $\widehat{Z}_{R}^{\eta _{BC}n}$ is a Gaussian noise with
variance $1-\frac{c_{2}^{2}}{c_{1}^{2}}$. From the conditional Entropy Power
Inequality, we now have
\end{subequations}
\begin{eqnarray}
2^{\frac{2}{\eta _{BC}n}h(Y_{R2}^{\eta _{BC}n}|Y_{1}^{n},W_{2})} &\geq &2^{%
\frac{2}{\eta _{BC}n}h\left( \frac{c_{2}}{c_{1}}Y_{R1}^{\eta
_{BC}n}|Y_{1}^{n},W_{2}\right) }  \label{epiYR2} \\
&&+2^{\frac{2}{\eta _{BC}n}h\left( \widehat{Z}_{R}^{\eta
_{BC}n}|Y_{1}^{n},W_{2}\right) }.  \notag
\end{eqnarray}%
Also, given that $a_{12}\leq 1$, we have,
\begin{subequations}
\begin{align}
h(Y_{R2}^{\eta _{BC}n}|Y_{1}^{n},W_{2})& =h(Y_{R2}^{\eta
_{BC}n}|X_{1}^{n}+Z_{1}^{n},W_{2})  \label{hYR2d} \\
& =h(Y_{R2}^{\eta _{BC}n}|X_{1}^{n}+Z_{2}^{n},W_{2}) \\
& \leq h(Y_{R2}^{\eta _{BC}n}|a_{12}X_{1}^{n}+Z_{2}^{n},W_{2}) \\
& =h(Y_{R2}^{\eta _{BC}n}|Y_{2}^{n},W_{2}) \\
& =\frac{\eta _{BC}n}{2}\log (2\pi e(1+c_{2}^{2}\xi P_{R}))
\end{align}%
due to the Markov chain $a_{12}X_{1}^{n}+Z_{2}^{n}\rightarrow
X_{1}^{n}+Z_{2}^{n},W_{2}\rightarrow Y_{R2}^{\eta _{BC}n}$. The proof is
concluded with standard steps.

\section*{B. Proof of Proposition 2 and 3}

The achievable region of Prop. 3 is obtained following Sec. \ref{sec_scenario}. $%
S_{1}$ transmits a private message $W_{1p}$ over the IC, $%
X_{1}^{n}(W_{1})=X_{1p}^{n}(W_{1p}),$ and an independent private message $%
W_{1R}$ over the OBRC via standard "Gaussian codebooks". $S_{2}$
transmits common messages $(W_{2c^{\prime }},W_{2c^{\prime \prime }})$ over
the IC ($X_{2}^{n}(W_{2})=X_{2c}^{n}(W_{2c^{\prime }},W_{2c^{\prime \prime
}})$), and an independent private message $W_{2R}$, along with $%
W_{2c^{\prime }}$ (interference forwarding), on the OBRC. Then, the
following conditions are easily seen to provide an achievable region
\vspace{-0.05in}
\end{subequations}
\begin{subequations}
\begin{align}
R_{1p}& \leq \mathcal{C}(P_{1}) \\
R_{2c^{\prime \prime }}+R_{1p}& \leq \mathcal{C}(P_{1}+a_{21}^{2}P_{2}) \\
R_{2c}& \leq \mathcal{C}\left( \frac{P_{2}}{1+a_{12}^{2}P_{1}}\right) \\
R_{1R}& \leq \eta _{MAC}\mathcal{C}(b_{1}^{2}P_{1R}) \\
R_{2c^{\prime }}+R_{2R}& \leq \eta _{MAC}\mathcal{C}(b_{2}^{2}P_{2R}) \\
R_{1R}+R_{2c^{\prime }}+R_{2R}& \leq \eta _{MAC}\mathcal{C}%
(b_{1}^{2}P_{1R}+b_{2}^{2}P_{2R}) \\
R_{2c^{\prime }}+R_{1R}& \leq \eta _{BC}\mathcal{C}(c_{1}^{2}\xi P_{R}) \\
R_{2R}& \leq \eta _{BC}\mathcal{C}\left( \frac{c_{2}^{2}\overline{\xi }P_{R}%
}{1+c_{2}^{2}\xi P_{R}}\right)
\end{align}%
Using Fourier-Motzkin elimination method, with the fact that $%
R_{1}=R_{1p}+R_{1R}$, $R_{2c}=R_{2c^{\prime }}+R_{2c^{\prime \prime }}$, and
$R_{2}=R_{2c}+R_{2R}$, the achievable region in Proposition 3 can be
obtained. Now, for $b_{2},c_{1}\rightarrow \infty $, the achievable region
becomes
\end{subequations}
\begin{eqnarray}  \label{prop14R2}
R_{1}&\leq& \mathcal{C}(P_{1})+\eta _{MAC}\mathcal{C}(b_{1}^{2}P_{1R})
\label{prop14R1} \\
R_{2}&\leq& \mathcal{C}\left( \frac{P_{2}}{1+a_{12}^{2}P_{1}}\right) +\eta
_{BC}\mathcal{C}\left( c_{2}^{2}P_{R}\right)
\end{eqnarray}%
since the overall region is maximized for $\xi =0$ for $b_{2},c_{1}%
\rightarrow \infty $. The converse of Prop. 3 is again obtained from\ Prop. 1,
similar to Prop. 2.

\end{document}